\documentclass[lettersize,journal]{IEEEtran}
\usepackage{amsmath,amsfonts}
\usepackage{amssymb}
\usepackage{algorithmic}
\usepackage{algorithm}
\usepackage{array}
\usepackage[caption=false,font=normalsize,labelfont=sf,textfont=sf]{subfig}
\usepackage[colorlinks=true, linkcolor=black]{hyperref}
\usepackage{textcomp}
\usepackage{stfloats}
\usepackage{url}
\usepackage[nocompress]{cite}
\usepackage{verbatim}
\usepackage{graphicx}
\usepackage{booktabs}  
\usepackage{multirow}  
\usepackage{makecell}
\usepackage{orcidlink} 
\begin{document}

\title{GCDance: Genre-Controlled Music-Driven \\ 3D Full Body Dance Generation}

\author{Xinran Liu~\orcidlink{0000-0002-8268-0286}, 
Xu Dong~\orcidlink{0009-0006-6842-8750}, 
Shenbin Qian~\orcidlink{0000-0002-6313-3359}, 
Diptesh Kanojia~\orcidlink{0000-0001-8814-0080}, \\
Wenwu Wang~\orcidlink{0000-0002-8393-5703},
~\IEEEmembership{Senior Member,~IEEE,} 
and Zhenhua Feng*~\orcidlink{0000-0002-4485-4249}
,~\IEEEmembership{Senior Member,~IEEE,}

\thanks{X. Liu and D. Knojia is with the School of Computer Science and Electronic Engineering,
University of Surrey, Guildford GU2 7XH, UK (e-mail: xinran.liu@surrey.ac.uk; d.kanojia@surrey.ac.uk)}
\thanks{X. Dong  is with the Department of Music and Media, University of Surrey, Guildford GU2 7XH, UK (e-mail: xu.dong@surrey.ac.uk)}
\thanks{S. Qian is with the Department of Informatics, University of Oslo, 0316 Oslo, Norway (email: shenbinq@ifi.uio.no)}
\thanks{W. Wang is with the Centre for Vision, Speech and Signal Processing, University of Surrey, Guildford GU2 7XH, UK (e-mail: w.wang@surrey.ac.uk).}      
\thanks{Z. Feng is with the School of Artificial Intelligence and Computer Science, Jiangnan University, Wuxi 214122, China (e-mail: fengzhenhua@jiangnan.edu.cn)}%
\thanks{*Corresponding Author}
}


\maketitle

\begin{abstract}
Music-driven dance generation is a challenging task as it requires strict adherence to genre-specific choreography while ensuring physically realistic and precisely synchronized dance sequences with the music's beats and rhythm. Although significant progress has been made in music-conditioned dance generation, most existing methods struggle to convey specific stylistic attributes in generated dance. To bridge this gap, we propose a diffusion-based framework for genre-specific 3D full-body dance generation, conditioned on both music and descriptive text. To effectively incorporate genre information, we develop a text-based control mechanism that maps input prompts, either explicit genre labels or free-form descriptive text, into genre-specific control signals, enabling precise and controllable text-guided generation of genre-consistent dance motions. Furthermore, to enhance the alignment between music and textual conditions, we leverage the features of a music foundation model, facilitating coherent and semantically aligned dance synthesis. Last, to balance the objectives of extracting text-genre information and maintaining high-quality generation results, we propose a novel multi-task optimization strategy. This effectively balances competing factors such as physical realism, spatial accuracy, and text classification, significantly improving the overall quality of the generated sequences. Extensive experimental results obtained on the FineDance and AIST++ datasets demonstrate the superiority of GCDance over the existing state-of-the-art approaches.
\end{abstract}


\section{Introduction}
\IEEEPARstart{D}{ancing} is a universal form of cultural expression and a powerful medium for conveying emotions. However, choreography is an artistic skill that demands years of training. During the choreographic process, the body movements of the choreographer need to be aligned with the musical rhythm while reflecting the stylistic characteristics of a specific dance genre~\cite{bannerman2014dance}. As a result, the use of AI for music-driven choreography shows promising research potential.

In recent years, numerous deep-learning-based approaches have been developed for the task of dance generation. 
Early dance generation approaches often rely on autoregressive models that directly predict future dance movements from the past motion sequences~\cite{li2021ai,kim2022brand}, but they frequently encounter challenges such as motion freezing during long-term generation. 
To mitigate this issue, Vector-Quantized Variational AutoEncoder (VQ-VAE) based methods~\cite{siyao2022bailando,gong2023tm2d} introduce a discrete codebook of motion units, effectively stabilizing long-range motion. 
Nevertheless, the reliance on a fixed latent vocabulary inherently restricts the diversity and expressiveness of generated dances~\cite{tevet2022humanmotiondiffusionmodel}. 
More recently, diffusion models~\cite{ho2020denoising} have shown remarkable performance in various generation tasks. Unlike methods that rely on predefined seeds or fixed latent vocabularies, diffusion models iteratively refine noise into coherent outputs, thereby capturing a broader space of potential motions. These approaches~\cite{tseng2023edge, qi2023diffdance, liu2024dgfm} greatly enhance both diversity and expressiveness of dance motions generated. 
However, existing approaches often struggle to convey specific stylistic attributes. Although these methods can generate a single style of dance for a given piece of music, they may lead to mismatches between the generated motion and the musical style, or may fail to produce dances that align with a user-intended genre.

\begin{figure}[!t]
  \centering
  \includegraphics[width=.5\textwidth]{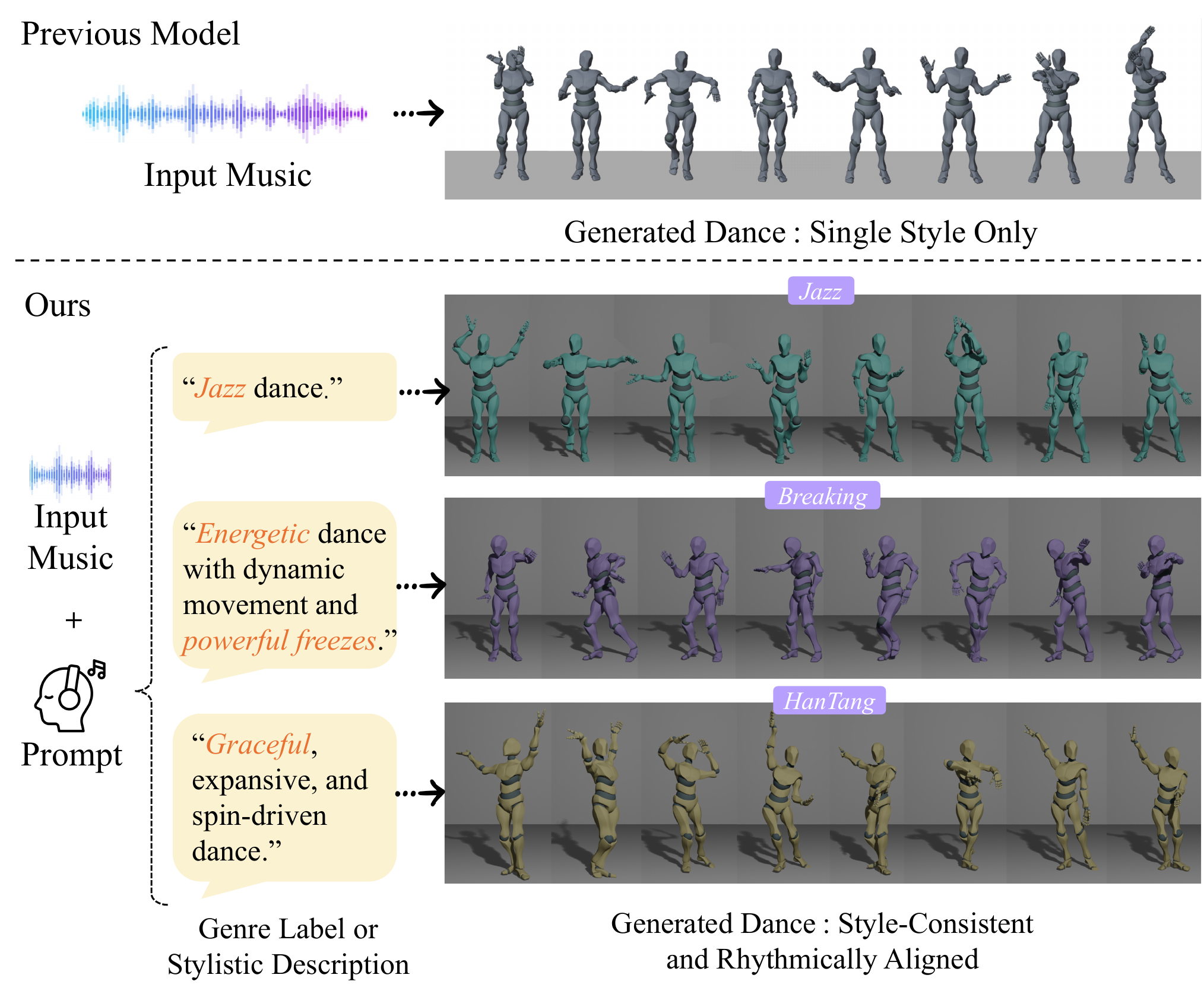}
  \caption{Given an audio input and a genre-descriptive textual prompt, GCDance generates 3D dance sequences that align well with the musical melody and beat while adhering to the textual instruction.}
  \vspace{-0.5cm}
\end{figure}

To address these limitations, we propose GCDance, a genre-controllable 3D full-body dance generation model conditioned on both music and text. GCDance focuses on generalization to high-fidelity motions while maintaining controllability. 
Specifically, we introduce a classification-based control mechanism utilizing explicit genre labels or descriptive natural language prompts as input. 
The textual input is first classified to determine its corresponding dance genre, and subsequently encoded into control signals to guide the generation process, enabling the model to modulate the generated dance style accordingly. 
With the introduction of the text as an additional conditioning modality, aligning it with the music representation is critical for achieving consistent and controllable dance generation. However, most existing dance generation methods rely solely on hand-crafted musical features~\cite{li2024lodge,yang2023longdancediff}, which are typically low-level and insufficient for capturing the complex and nuanced correlations between music and textual descriptions. To achieve a better alignment between these multimodal signals, we integrate hand-crafted features with deep features obtained from the Wav2CLIP music foundation model~\cite{wu2022wav2clip}. Since Wav2CLIP projects audio and text into a shared embedding space, this alignment leads to a unified understanding of the musical and genre characteristics that drive dance movements, which in turn helps the model generate stylized dance motions conditioned on diverse textual prompts.

Apart from the above genre-controlled mechanism, achieving robust dance generation inherently involves multiple objectives.
This requires the model to balance goals such as spatial accuracy, temporal coherence, and genre control. 
In practice, these objectives may conflict with each other, leading to trade-offs in the generated motions. For example, increasing motion diversity can reduce fidelity and coherence, and highly realistic sequences may still fail to reflect the intended genre style. 
Existing approaches typically consolidate these competing objectives into a single loss function with manually tuned weights~\cite{tseng2023edge,li2023finedance}, often leading to suboptimal trade-offs among different aspects of motion quality. 
To achieve a dynamic balance among these tasks and enhance the performance of generated dances, we adopt a Multi-Task Learning (MTL) framework that jointly optimizes multiple objectives such as motion quality, velocity constraints, foot contact consistency, and genre classification. By assigning a distinct objective function to each requirement, our method dynamically adjusts the training process and ultimately improves motion quality, achieving state-of-the-art performance across multiple quantitative evaluation metrics.
In addition, our model is trained on a dataset with 52-joint full-body representations, which include detailed hand movements. 
This richer skeletal representation further enhances the realism and expressiveness of the generated dances by capturing fine-grained motion details that are neglected in previous studies.

In summary, the main contributions of GCDance include:
\begin{itemize}
    \item We introduce a diffusion-based multi-genre dance generation model, namely GCDance. It enables controllable dance generation by conditioning on both music and textual prompts. 
    
    

    \item To enhance cross-modal alignment, GCDance leverages a pretrained music foundation model that captures both high-level semantic cues and low-level audio details for more coherent and expressive dance generation.

    \item We introduce a novel multi-task learning framework that jointly optimizes diverse objectives for a more balanced model training.
    
    \item Extensive experimental results obtained on both the FineDance and AIST++ datasets demonstrate the superiority of the proposed GCDance method over the existing approaches.
\end{itemize}

\section{Related work}
\subsection{Music Driven Dance Generation}
Early studies~\cite{ofli2008audio, fukayama2015music} approach this task as a similarity-based retrieval problem, where motion segments are selected from a predefined database based on the input music. These methods inherently limit the diversity and creativity of the generated dances. To overcome these limitations, deep learning models reframe the task as motion prediction using architectures such as Convolutional Neural Network (CNN)~\cite{holden2016deep}, Recurrent Neural Network (RNN)~\cite{chiu2019action,du2019bio}, and Transformers~\cite{fan2022bi,huang2022genre,li2022danceformer}. 
However, these frame-by-frame prediction approaches often face challenges such as error accumulation and motion freezing~\cite{zhuang2022music2dance}. 

Recent research has shifted to the generative pipeline. Based on VQ-VAE, TM2D~\cite{gong2023tm2d} incorporates music and text instructions to generate coherent dance segments with the given music while retaining semantic information. Bailando~\cite{siyao2022bailando} quantizes meaningful dance units into a quantized codebook and employs a reinforcement-learning-based evaluator to improve the alignment between generated movements and musical beats. Despite their outstanding performance, these systems are highly complex and involve multiple sub-networks. 
EDGE~\cite{tseng2023edge} is the first method that employs a diffusion model for dance generation, featuring a single-model design optimized for a single objective. However, existing models are typically trained on datasets containing only $24$ body joints and overlook the quality of hand motion generation. To address this limitation, Li~\textit{et al.}~\cite{li2023finedance} proposed FineNet and introduced a new dataset with $52$ joints. 
It is also worth mentioning that the vast majority of models rely on handcrafted musical features such as Mel-Frequency Cepstral Coefficient (MFCC), chroma, or one-hot beat features, which may not fully capture intricate details needed for fine-grained dance movement correlation.


\subsection{Diffusion Models}
Diffusion models~\cite{ho2020denoising,nichol2021improved} are a type of deep generative model and have made significant progress in recent years~\cite{croitoru2023diffusion}.
They have been widely applied across multiple fields of research, such as image generation~\cite{ruiz2023dreambooth,ramesh2022hierarchical}, audio synthesis~\cite{liu2024audioldm2,liu2023audioldm,kong2020diffwave}
and text generation~\cite{lovelace2024latent,he2022diffusionbert}. For conditional generation, existing approaches often employ classifier guidance~\cite{chung2022improving,dhariwal2021diffusion} or classifier-free guidance~\cite{nichol2021glide,rombach2022high} to enhance the quality of sample generation, which is applicable to any pretrained diffusion model to improve performance without retraining. 
Furthermore, the growing interest in diffusion models is attributed to their remarkable ability for controllable generation. Blended Diffusion~\cite{avrahami2022blended} presents a text-conditional image generation model, utilizing CLIP~\cite{radford2021learning} to guide the diffusion process to produce images that conform to the target prompt. GMD~\cite{karunratanakul2023guided} applies diffusion to the task of text-to-motion trajectory generation, integrating spatial constraints to improve the alignment between spatial information and local poses. Alexanderson \textit{et al.}~\cite{alexanderson2023listen} propose an audio-driven motion generation focusing on gestures and dance, and also implement style control and strength adjustment of stylistic expression. However, this method is limited to only four genres. Dance motion generation is a more complex task and suffers from lower data availability due to its specialized nature~\cite{tseng2023edge}. In our work, we present a diffusion-based method that can not only generate 16 different dance genres conditioned on music, but also control the type of dance through textual prompts.
\begin{figure*}[!t]
  \centering
  \includegraphics[width=.955\textwidth]{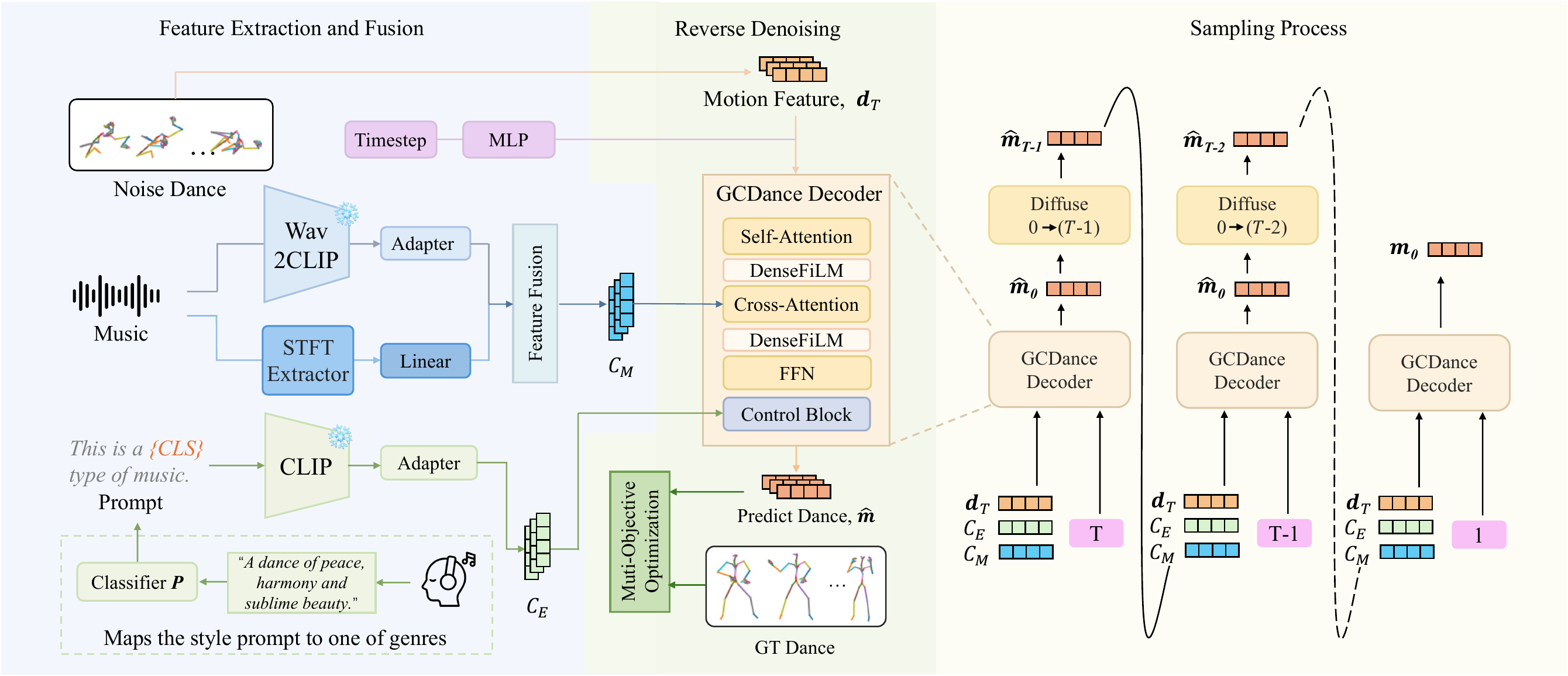}
  \caption{An overview of GCDance. Left: the multimodal inputs and feature extraction. Middle: the training process at a given diffusion timestep $t$. Right: the sampling process, where a sequence of dance motions is generated iteratively.}
  \vspace{-0.5cm}
  \label{f1}
\end{figure*}
\subsection{Multi-Task Learning}
Multi-Task learning (MTL) trains related tasks simultaneously using a shared representation. Although early MTL methods sometimes underperform single-task models~\cite{Standley2020}, recent approaches have overcome these issues. For example, MTAN~\cite{liu2019end} is a multi-task learning architecture that uses dynamic weight averaging with task-specific feature-level attention by employing a shared network and soft-attention modules without preset weighting schemes. Similarly, an impartial MTL was proposed in~\cite{liu2021towards}, which uses distinct strategies for shared and task-specific parameters. In addition, Nash-MTL~\cite{pmlr-v162-navon22a} re-frames the gradient combination as a bargaining game, using the Nash Bargaining Solution~\cite{Nash1953} to negotiate a joint update direction among tasks. To improve training stability, Aligned MTL was developed~\cite{Senushkin2023}, which aligns the orthogonal components of gradient systems according to their condition number. Furthermore, a Bayesian gradient aggregation method was introduced to model uncertainty over task-specific parameters and gradients~\cite{pmlr-v235-achituve24a}. These advances have been widely applied in various fields in computer vision~\cite{Zhao_2018_ECCV, Wong_2023_CVPR} and natural language processing~\cite{deoghare-etal-2023-multi,qian-etal-2024-multi}.

Combining different training objectives is common in dance generation. However, existing approaches typically consolidate these
competing objectives into a single loss function with manually tuned weights~\cite{tseng2023edge,li2023finedance}, rather than weights learned by parametric heuristics. This often leads to suboptimal trade-offs
among different aspects of motion quality. To address this issue, we propose a novel multi-objective training strategy that integrates parametric loss heuristics like Nash MTL and Aligned MTL to optimize these training objectives.


\section{The Proposed GCDance Method}
In this section, we present the details of the proposed GCDance method, introducing its overall architecture and key components. The diffusion preliminaries of our approach are provided in the supplementary material.

\subsection{The GCDance Architecture}

The overall architecture of GCDance is illustrated in Figure~\ref{f1}. We define three modalities in the framework: dance motion, music, and textual prompt. Each modality is turned into an informative representation as detailed below.

Given a long music-dance pair, we first divide it into $N$ $4$-second segments. For each segment, we uniformly sample $k$ frames from the corresponding dance motion and music clip. 

For the \textbf{dance motion representation}, according to the Skinned Multi-Person Linear (SMPL) format~\cite{loper2023smpl}, we define three components. ($1$) Human joint positions: We transform the $52$ joint positions into a $312$ (i.e. $52\times 6$) dimensional rotation representation with $6$ degrees of freedom (DOF), denoted as $\textbf{\textit{p}} \in \mathbb{R}^{312}$.
($2$) Root translation: A $3$D vector is used to describe the global translation of the root joint. ($3$) Foot-ground contact: Following the approach in EDGE~\cite{tseng2023edge}, we incorporate a $4$D foot-ground contact label to represent the binary states of heel and toe ground contact for each foot, given by $\textbf{\textit{f}} \in \mathbb{R}^{4}$. Consequently, the complete representation of the pose sequence is $\textbf{\textit{m}} \in \mathbb{R}^{k\times 319}$, where $k$ represents the number of frames.

\begin{figure}[!t]
  \centering
  \includegraphics[width=.41\textwidth]{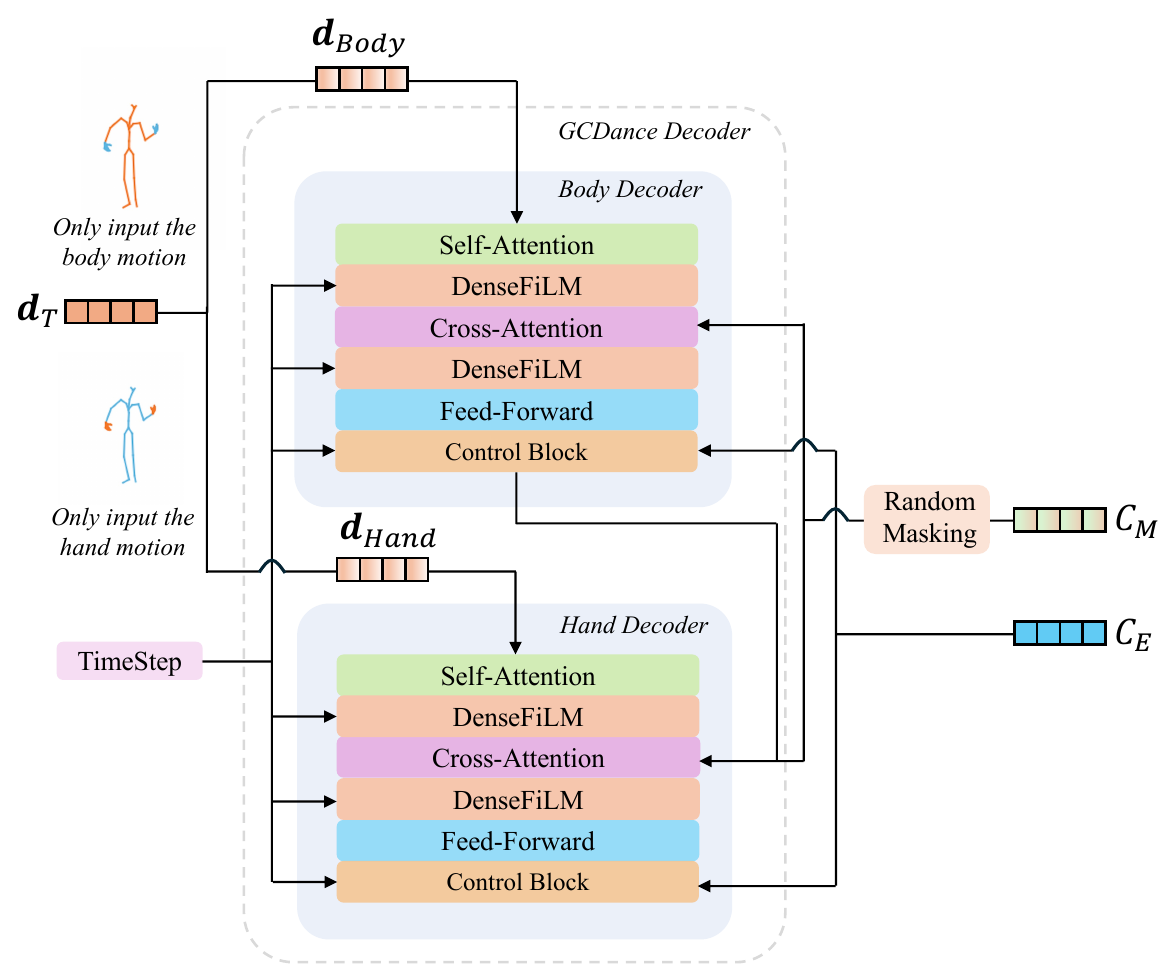}
  \caption{The decoder of GCDance.}
  \vspace{-0.5cm}
  \label{f2}
\end{figure}

For \textbf{music representations}, existing approaches typically rely on hand-crafted musical features, overlooking recent advances in music foundation models, which have shown strong potential for capturing nuanced representations of music. To address this limitation, GCDance integrates music embeddings extracted from a pretrained music foundation model with hand-crafted music features, effectively leveraging the advantages of high-level semantic information and low-level temporal details to improve the quality of the generated dance sequences. For high-level representations, we adopt Wav2CLIP~\cite{wu2022wav2clip} as the music encoder. Wav2CLIP is an audio-visual correspondence model that distills from the CLIP framework~\cite{radford2021learning}. It is trained to predict CLIP-style embeddings from raw audio by aligning them with frozen vision-based representations extracted from videos. For hand-crafted music features, we employ Short-time Fourier Transform (STFT) that captures fine-grained temporal-frequency features in music signals~\cite{gourisaria2024comparative}. 
In GCDance, we extract STFT features using the Librosa toolbox~\cite{mcfee2015librosa}.

For \textbf{text representations}, our goal is to establish a free form text guided dance generation framework. However, the absence of text–dance paired data in existing datasets presents a significant challenge. To address this issue, we construct a dance genre description dataset and develop a genre classifier $\textit{\textbf{P}}$ that maps free-form textual descriptions $C_{\text{desc}}$ to genre $\hat{g}$. Comprehensive details of this dataset are provided in the supplementary materials.

To evaluate the performance of the genre classifier, we compute the binary cross entropy (BCE) loss between the predicted distribution $\hat{g}$ and the ground truth genre label $g$ associated with the input music–text pair:
\begin{equation} 
\hat{g} = P(C_{\text{desc}}),
\end{equation}  
\begin{equation} 
L_{C} = BCE\left ( \hat{g},g \right ) 
\end{equation}  

Based on the predicted genre $\hat{g}$, we apply a prompt learning strategy~\cite{zhou2022learning} to transform the discrete label into a complete textual prompt, thereby providing genre-related semantic information to guide the generation process. For example, given the genre label ``Jazz," the generated sentence is  ``This is a Jazz type of music.".
CLIP is then employed to encode this prompt into a semantic embedding, denoted as $C_E$, which captures genre-specific textual semantics aligned with the user’s input.

Finally, GCDance takes as input the noise slice $\textbf{\textit{d}}_{T}$, the music condition $C_{M}$, the text genre embedding $C_{E}$, and the diffusion timestep $t$. These inputs are then fed into a Transformer-based denoising network. 
As illustrated in Figure~\ref{f2}, we employ two expert downsampling modules to separately model the distributions of body motion and hand motion inspired by~\cite{li2023finedance}. This approach is motivated by the distinctions in the range of motion and degrees of freedom between the body and hands. By learning their unique feature spaces independently, the model can generate dance sequences with enhanced detail and expressiveness. 

To elaborate on the process, the motion sequences are separately fed into the two Transformer-based networks. They consist of a self-attention module, a cross-attention module, multilayer perceptrons, and Feature-wise Linear Modulation (FiLM) layers~\cite{perez2018film}.  
The output features from the body decoder are integrated into the cross-attention layer of the hand decoder to help capture the relationship between body and hand movements effectively. However, a significant domain gap still exists between the raw conditional features and the dance motions. To bridge this gap, we introduce an adapter module to process the extracted music and text representations and effectively align them in the latent space. Additionally, to incorporate the music conditioning input, we utilize a cross-attention mechanism to process music features projected into the embedding space.

\begin{figure}[!t]
  \centering
  \includegraphics[width=.27\textwidth]{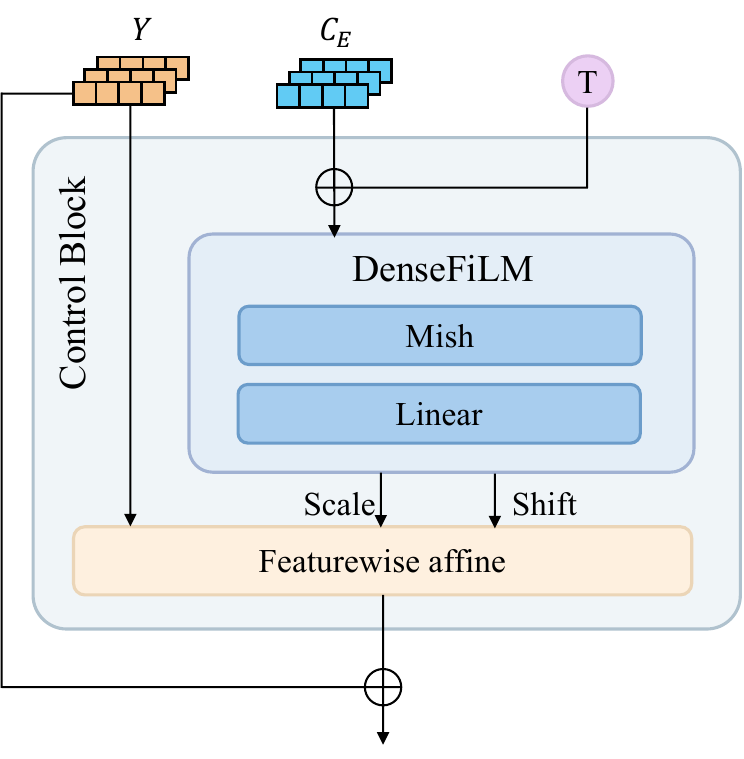}
  \caption{The control module of GCDance.}
  \vspace{-0.5cm}
  \label{f2.5}
\end{figure}

\textbf{Genre-Controllability.} As illustrated in Figure~\ref{f2.5}, the control module integrates genre information into the generation process at each diffusion timestep through a FiLM layer. 
FiLM modulates the intermediate activations of the network through affine transformations conditioned on external inputs, enabling dynamic adaptation of the representation based on contextual signals.

In GCDance, we use the output from the previous network layer, denoted as $Y$, along with a genre embedding $C_{E}$ as inputs to the control module. 
The genre embedding is conditioned on the current diffusion timestep and then used to derive the FiLM modulation parameters as follows:

\begin{equation}  
\quad\gamma  =\theta_{w} (\alpha(C_{E})), \quad\varepsilon  = \theta_{b} (\alpha(C_{E})) 
\end{equation}  
\begin{equation}  
FiLM_{t}(Y) = \gamma \odot Y + \varepsilon,
\end{equation}  
where $\alpha$ is a text embedding adapter used to adjust the embedding representation, $\odot$ denotes element-wise multiplication, and $\theta_{w}$ and $\theta_{b}$ are learned linear projections.

\subsection{Multi-Objective Training} 
\textbf{Training Objective}. The training process involves five objectives. We adopt the loss function $\mathcal{L}_{S}$ from DDPM as the primary objective, which is defined as:
\begin{equation}  
\label{equ_2}
\mathcal{L}_{\text {S}}=
\mathbb{E}_{\boldsymbol{ \textbf{\textit{m}}_{0}}, t}\left[\left\| \textit{\textbf{m}}_{0}- f_{rev} (\textit{\textbf{d}}_{t}, t, 
C_{M}, C_{E}  )\right\|_{2}^{2}\right]
\end{equation}

In addition, to generate fluent and physically-plausible motion sequences, we incorporate several auxiliary losses frequently used in motion generation tasks, such as EDGE~\cite{tseng2023edge} and Motion Diffusion Model (MDM)~\cite{tevet2022humanmotiondiffusionmodel}. 
These auxiliary losses encourage alignment in three key aspects: joint positions (Equ.~\ref{eq:4-3}), velocities (Equ.~\ref{eq:4-4}), and foot contact (Equ.~\ref{eq:4-5}). 
Similar to previous studies~\cite{tevet2022humanmotiondiffusionmodel}, we use the forward kinematic function $FK(\cdot)$ to transform the joint angles into their corresponding joint positions, calculating the joint loss:
\begin{equation}  
\mathcal{L}_{\text {J}}=\frac{1}{k} \sum_{j=1}^{k}\left\|F K\left(\boldsymbol{\textbf{\textit{m}}}^{j}\right)-F K\left(\hat{\boldsymbol{\textbf{\textit{m}}}}^{j}\right)\right\|_{2}^{2}
\label{eq:4-3}
\end{equation}  
where $j$ represents the frame index and $\hat{\boldsymbol{\textbf{\textit{m}}}}^{j}$ represents the predicted pose for this frame.
We also compute velocity and acceleration, introducing the velocity loss:
\begin{equation}  
\mathcal{L}_{\text {V}}=\frac{1}{k-1} \sum_{j=1}^{k-1}\left\|\left(\boldsymbol{\textbf{\textit{m}}}^{j+1}-\boldsymbol{\textbf{\textit{m}}}^{j}\right)-\left(\hat{\boldsymbol{\textbf{\textit{m}}}}^{j+1}-\hat{\boldsymbol{\textbf{\textit{m}}}}^{j}\right)\right\|_{2}^{2}
\label{eq:4-4}
\end{equation}  
Lastly, we apply the contact loss $\mathcal{L}_{\text {F}}$ that leverages binary foot-ground contact labels to optimize the consistency in foot contact during motion generation:
\begin{equation}
\mathcal{L}_{\text {F}}=\frac{1}{k-1} \sum_{j=1}^{k-1}\left\| \left( F K\left(\hat{\boldsymbol{\textbf{\textit{m}}}}^{j+1}\right)-F K\left(\hat{\boldsymbol{\textbf{\textit{m}}}}^{j}\right) \right) \cdot \hat{\textbf{\textit{b}}}^{j} \right\|_{2}^{2}
\label{eq:4-5}
\end{equation}
where $\hat{\textbf{\textit{b}}}^{j}$ is the predicted binary foot-ground contact label. 

To balance multiple training objectives and address the optimization challenges such as conflicting or dominating gradients, we propose a multi-objective training strategy below:
\begin{equation}
\mathcal{L}=\tau \left ( \mathcal{L}_{\mathrm{S}}, \mathcal{L}_{\mathrm{J}}, \mathcal{L}_{\mathrm{V}},
\mathcal{L}_{\mathrm{F}},
\mathcal{L}_{\mathrm{C}}, \right ) 
\label{eq:4-6}
\end{equation}

This strategy relies on a heuristic function $\tau$ that combines five distinct losses into a single optimization objective. The goal is to find a parameter set $\theta$ that minimizes the overall aggregation loss:
\begin{equation}
\Delta\theta = \min_{\theta}\sum_{i=1}^{T} \mathcal{L}_{i}\left ( \theta_{i}  \right ) 
\label{eq:4-7}
\end{equation}
where $T$ denotes the number of loss components, and $\mathcal{L}_i(\theta_i)$ represents the $i$-th loss function.

\textbf{MTL Training Strategy}. In our implementation, we explore two different heuristics, including Nash MTL~\cite{pmlr-v162-navon22a} and Aligned MTL~\cite{Senushkin2023}, to learn the parameter set $\theta$. 

Nash MTL is designed to compute an update vector $\Delta\theta$ that integrates the task-specific gradients $g_i$, while ensuring that $\Delta\theta$ remains within an $\epsilon$-radius ball centered at zero, denoted by $B_\epsilon$. This is formulated as the following optimization problem:
\begin{equation} 
arg\ max_{\Delta\theta\in B_{\epsilon}} \Sigma_i log(\Delta\theta^\intercal g_i)
\label{eq:nash}    
\end{equation}
The optimal solution to this problem is (up to scaling) $\Sigma_i \alpha_i g_i$, where $\alpha \in \mathbb{R}_+^K$ is the solution to $G^\intercal G \alpha = 1 / \alpha$ with the reciprocal taken element-wise. The complete Nash MTL algorithm is outlined below:
\begin{algorithm}[H]
\caption{Nash-MTL}
\label{alg:nash-mtl}
{\normalsize
\begin{algorithmic}[1]
\REQUIRE Initial parameter vector $\theta^{(0)}$, differentiable loss functions $\{l_i\}_{i=1}^K$, learning rate $\eta$
\FOR{$t = 1, \dots, T$}
    \STATE Compute task gradients $g_i^{(t)} \leftarrow \nabla_{\theta^{(t-1)}} l_i$
    \STATE Form matrix $G^{(t)}$ with columns $g_i^{(t)}$
    \STATE Solve for $\alpha$: $(G^{(t)})^\intercal G^{(t)} \alpha = 1 / \alpha$ to obtain $\alpha^{(t)}$
    \STATE Update parameters: $\theta^{(t)} \leftarrow \theta^{(t-1)} - \eta G^{(t)} \alpha^{(t)}$
\ENDFOR
\RETURN $\theta^{(T)}$
\end{algorithmic}
}
\end{algorithm}
\vspace{-0.3cm}

Aligned MTL is a method that aligns the principal components of the gradient matrix to enhance training stability. As formulated in Equ.~\ref{eq:aligned-obj}, it aims to reduce the discrepancy between the original gradient matrix $G$ and its aligned version $\hat{G}$, with the difference measured using the Frobenius norm. Moreover, the constraint in Equ.~\ref{eq:aligned-obj} mandates that $\hat{G}$ be orthogonal, meaning that its transpose multiplied by itself is equal to the identity matrix. This orthogonality condition is crucial for ensuring stability in the gradient's linear system.
\begin{equation} 
\underset{\hat{G}}{min} \lVert G- \hat{G} \rVert^2_F \ \ s.t.\ \ \hat{G}^\intercal \hat{G} = I
\label{eq:aligned-obj}
\end{equation}
\begin{equation} 
\hat{G} = \sigma UV^\intercal = \sigma GV \Sigma^{-1}V^\intercal
\label{eq:aligned-algo}
\end{equation}
Equ.~\ref{eq:aligned-algo} outlines the approach, where $\hat{G}$ is determined through singular value decomposition (SVD). In this procedure, the matrix G is factorized into three components: $U$, $\Sigma$, and $V^\intercal$. Here, both $U$ and $V$ are orthogonal matrices, while $\Sigma$ is a diagonal matrix that contains the singular values of $G$. The complete Aligned MTL algorithm is detailed below:

\begin{algorithm}[H]
\caption{Aligned-MTL}
\label{alg:aligned-mtl}
{\normalsize
\begin{algorithmic}[1]
\REQUIRE Gradient matrix $G \in \mathbb{R}^{|\theta| \times T}$, task importance $w \in \mathbb{R}^T$
\STATE Compute $M \leftarrow G^\intercal G$
\STATE Perform eigen-decomposition on $M$ : $(\lambda, V) \leftarrow \text{eigh}(M)$
\STATE Construct inverse root $\Sigma^{-1} \leftarrow \text{diag}\left(\sqrt{\tfrac{1}{\lambda_1}}, \dots, \sqrt{\tfrac{1}{\lambda_R}}\right)$
\STATE Compute transformation matrix $B \leftarrow \sqrt{\lambda_R} \cdot V \Sigma^{-1} V^\intercal$
\STATE Compute task weight vector $\alpha \leftarrow B w$
\RETURN $G \alpha$
\end{algorithmic}
}
\end{algorithm}
\vspace{-0.3cm}

\subsection{Sampling}
The sampling process is shown in the right part of Figure~\ref{f1}. At each denoising timestep $t$, unlike conventional generation models based on diffusion, which reconstruct the output by predicting the noise term $d_{t}$, our model directly predicts the dance pose $\hat{m}$. The predicted pose is then re-noised back to timestep $t-1$ as illustrated in Equ.~\ref{eq:4-2}.
\begin{equation}
\textit{\textbf{d}}_{t-1}\sim q(\hat{\textit{\textbf{m}}}(\textit{\textbf{d}}_{t},C_{E},C_{M}),t-1) 
\label{eq:4-2}
\end{equation}
This process is repeated until $t$ reaches zero. 

\textbf{Editing Sampling.}
Building on the previous method~\cite{tseng2023edge}, our approach enhances diversity by incorporating diffusion inpainting techniques.  In practice, our model allows users to apply a wide range of constraints.  Users can specify conditions for generating in-between movements in the temporal domain or for editing specific joint parts in the spatial domain. Based on these defined constraints, our method generates tailored dance outcomes, offering fine-grained control over the generated dance sequences. This process occurs only during sampling and is not included in the training process. For detailed mathematical formulations and implementation specifics, including examples of editing based on joint-wise or temporal constraints, please refer to the supplementary material.

\begin{figure*}[t!]
  \centering
  \includegraphics[trim={0 1cm 0 0}, clip, width=0.95\textwidth]{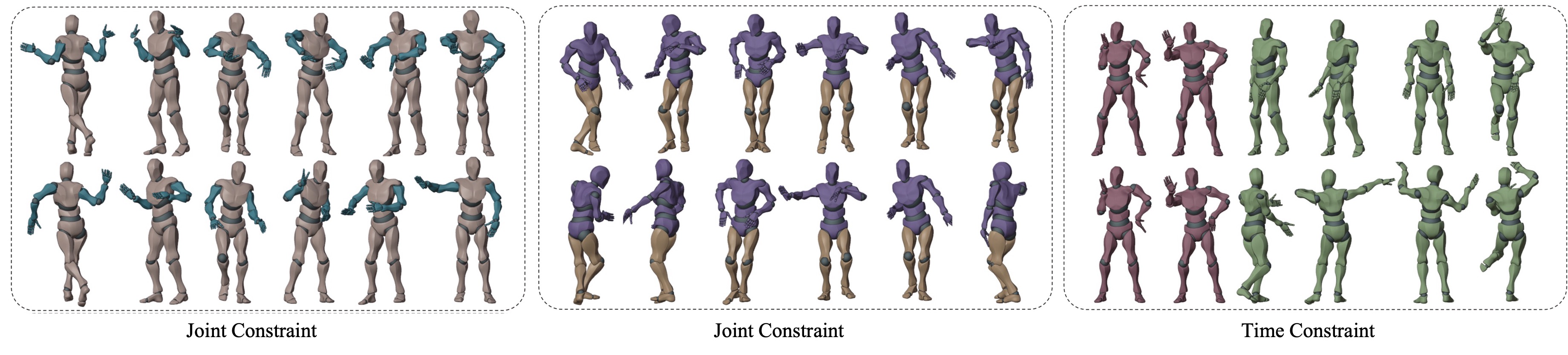}
  \caption{GCDance can generate joint-specific and temporally-specific dance segments. In the left example, the constrained body joints are shown in gray, while the generated hand joints are depicted in cyan. In the middle example, the constrained upper-body joints are shown in purple, and the generated leg joints are depicted in yellow. In the right example, the constrained first second is shown in red, while the generated last three seconds are depicted in green.}
  \vspace{-0.5cm}
  \label{edit}
\end{figure*}

To enable flexible editing of dance sequences, our method applies a diffusion inpainting mechanism during the sampling process, which allows users to apply a wide range of
constraints as shown in Figure~\ref{edit}.
Given a subset of joint-wise or temporal constraint inputs 
$ \textbf{\textit{m}}^{\textit{known}} $, with positions indicated by a binary mask $\textit{B}$, the model performs the following denoising steps during sampling. 
\begin{equation}
\textit{\textbf{d}}_{t-1} := \textit{B} \odot q(\textbf{\textit{m}}^{\textit{known}}, t-1) + (1 - \textit{B}) \odot \textit{\textbf{d}}_{t-1}
\end{equation}
where $\odot $ denotes the Hadamard product, an element-wise operation that substitutes the known part of the motion with noisy samples based on the specified constraint.

Taking the example of editing  dance sequence based on key joints, if we want to generate suitable hand joint motion
based on body movements. User can provide a reference motion $ \textbf{\textit{m}}^{\textit{known}} \in \mathbb{R}^{k \times 319} $ along with a mask $\textit{B}\in \left \{0,1\right \}^{k\times 319}$, where $\textit{B}$ has all $0$ for the hand joint features and all $1$ for the body joint features. 
This setup will generate a sequence of $k$ frames, where the body joint movements are based on the user-provided reference, and the hand joint regions are filled with consistent and coherent hand dance movements. The editing framework serves as a robust tool for downstream applications, offering flexible control over both temporal and spatial elements to create dance sequences that precisely conform to a variety of user-defined constraints.

\textbf{Long-term Sampling.}
Building on editing capability, our model further supports the generation of long-term dance sequences with temporal consistency. Specifically, given a long music sequence, we divide it into $N$ sub-sequences of $4$ seconds each. During the sampling process, GCDance constrains the first 2 seconds of each sequence to match the last $2$ seconds of the previous sequence. To further maintain consistency between adjacent $2$-second generated slices, we apply interpolation with linearly decaying weights to enhance performance. Through this approach, although our model is trained on 4-second clips, it can still synthesize dance sequences of any length by applying temporal constraints across batches of sequences. 

\section{Experimental Results \& Analysis}
In this section, we present the dataset, evaluation metrics, and comprehensive experimental results. Additional implementation details are provided in the supplementary materials.

\subsection{Dataset}
We evaluated the proposed method on the FineDance dataset~\cite{li2023finedance}, which contains $7.7$ hours of paired music and dance, totaling $831,600$ frames at $30$ frames per second (FPS) across $16$ different genres. The average dance length is $152.3$ seconds. 
The skeletal data of FineDance is stored in a 3D space and is represented by the standard $52$ joints, including the finger joints.
We trained all the methods on the $183$ pieces of music from the training set and generated $270$ dance clips across $18$ songs from the test set, using the corresponding real dances as ground truth.

We also conducted experiments on the widely used music-dance paired dataset AIST++~\cite{li2021ai}, which contains $1,363$ $3$D dance sequences paired with music, totaling $5.2$ hours of motion data across $10$ distinct genres at a frame rate of $60$ FPS. The dataset is constructed from multi-view dance videos and adopts a $24$-joint skeleton representation based on the SMPL model. We followed the experimental setting of Bailando~\cite{siyao2022bailando} for evaluation.

\begin{table*}[!t]
\caption{Comparison on the FineDance dataset. We use \textbf{bold} and \underline{underline} to highlight the best and second-best results. $\downarrow$ indicates that lower values are better, and vice versa for $\uparrow$. $*$ denotes abnormally high Div values caused by discontinuous motions~\cite{li2021ai}.
\label{table_1}}
\centering
\footnotesize
\resizebox{0.95\textwidth}{!}{%
\begin{tabular}{cccccclc}
\toprule
& \multicolumn{2}{c}{\textbf{Motion Quality}} & \multicolumn{2}{c}{\textbf{Motion Diversity}} & \multirow{2}{*}{\textbf{PFC}$\downarrow$}  &\multirow{2}{*}{\textbf{PBC}$\rightarrow$  }& \multirow{2}{*}{\textbf{BAS}$\uparrow$} \\  
\cmidrule(lr{0.35em}){2-3} \cmidrule(lr{0.35em}){4-5}
& FID\_hand$\downarrow$ & FID\_body$\downarrow$ & Div\_hand$\uparrow$ & Div\_body$\uparrow$ &  && \\ 
 \midrule
 GT &  /& /& 11.8156 ± 0.1314& 10.1810 ± 0.1327 & /&5.23 ± 0.16& 0.2318 ± 0.0070  \\
\midrule
DanceRevolution~\cite{huang2020dance}      & 219.52 ± 18.32    & 99.83 ± 7.79      & 1.85 ± 0.60        & 4.49 ± 0.25         & 6.81 ± 0.81      &23.39 ± 2.03& 0.2104 ± 0.0057       \\
MNET~\cite{kim2022brand}                  & 195.56 ± 5.04     & 154.79 ± 2.80       & 6.79 ± 0.20     & 8.25 ± 0.39\textsuperscript{*}     & 2.98 ± 0.11        &12.21 ± 0.15 & 0.1792 ± 0.0014       \\
Bailando~\cite{siyao2022bailando}          & 55.60 ± 8.15      & 57.77 ± 6.01         & 6.40 ± 0.68     & 4.27 ± 0.43        & 0.34 ± 0.01        &3.09 ± 0.06 & 0.2152 ± 0.0028       \\
EDGE~\cite{tseng2023edge}                  & 25.37 ± 3.24     & 51.56 ± 3.62           & 8.29 ± 0.30   &  5.88 ± 0.32         & 0.21 ± 0.03        &7.78 ± 0.07&  0.2171 ± 0.0056\\
FineNet~\cite{li2023finedance}             & 26.88 ± 3.09&  23.59 ± 3.56& 8.30 ± 0.45& 6.64 ± 0.28 & \textbf{0.12 ± 0.01} &3.35 ± 0.11& 0.2066 ± 0.0046\\
 DGFM~\cite{liu2024dgfm}& 20.699 ± 3.52& 24.63 ± 3.14& 8.77 ± 0.41& \underline{6.77 ± 0.75}& 0.20 ± 0.01& 4.23 ± 0.06&0.2153 ± 0.0054\\
 LODGE~\cite{li2024lodge} & 18.36 ± 2.10& 47.56 ± 1.37& 8.57 ± 0.36& 5.41 ± 0.27& \underline{0.13 ± 0.01}  & 3.46 ± 0.06&\textbf{0.2327 ± 0.0050}\\
\midrule
GCDance (Nash)& \textbf{17.69 ± 2.70} & \underline{22.90 ± 2.45} & \textbf{9.47 ± 0.38} & 6.39 ± 0.23 &  \underline{0.13 ± 0.01}  &\textbf{4.71 ± 0.06}& \underline{0.2238 ± 0.0056}\\

 GCDance (Aligned)& \underline{18.06 ± 3.12}& \textbf{21.67 ± 2.41} & \underline{9.01 ± 0.47} & \textbf{6.84 ± 0.75} &  0.15 ± 0.01  &\underline{4.54 ± 0.07}& 0.2205 ± 0.0041\\
\bottomrule
\end{tabular}}%
\vspace{-0.5cm}
\end{table*}


\subsection{Evaluation Metrics} 
We evaluated our approach based on four aspects: motion quality, generation diversity, motion-music correlation, and physical plausibility.

\textbf{Motion Quality}: Following the previous approaches~\cite{li2021ai}, we evaluate the motion quality using Fréchet Inception Distance (FID)~\cite{heusel2017gans}. This metric measures the dissimilarity between the feature distributions of generated dance sequences and ground truth dance sequences by computing the distribution distance in the feature space.

\textbf{Generation Diversity}: We follow Bailando~\cite{siyao2022bailando} and quantify diversity by calculating the average Euclidean distance of kinetic features across the generated motions.

\textbf{Motion-Music Correlation}: 
To evaluate the alignment between music beats and motion transition beats, we employ the Beat Alignment Score (BAS)~\cite{siyao2022bailando} metric, which assesses the correlation between motion and music by calculating the average temporal distance between each kinematic beat and its nearest musical beat. 

\textbf{Physical Plausibility}: 
We adopt the Physical Foot Contact (PFC) metric~\cite{tseng2023edge}, which is inspired by the principles of center-of-mass (COM) motion and its relationship with foot-ground contact, and the Physical Body Contact (PBC) score~\cite{luo2024popdg}, an extension of PFC that further accounts for upper-body dynamics by incorporating signals from the neck and hands.

\subsection{Quantitative Results}

In Table~\ref{table_1}, we compared our method with recent methods: DanceRevolution~\cite{huang2020dance}, MNET~\cite{Kim_2022_CVPR}, Bailando~\cite{siyao2022bailando}, EDGE~\cite{tseng2023edge}, FineNet~\cite{li2023finedance}, DGFM~\cite{liu2024dgfm} and LODGE~\cite{li2024lodge}.
Among these, only FineNet was originally trained on the FineDance dataset. For a fair comparison, we retrained the other methods on FineDance using their publicly available code and default training configurations. 
MNET is the only baseline that also incorporates genre information during generation. 
For each model, we generated $10$ sets of dance sequences, with each set randomly sampled from $270$ dance clips in the test set. 
Each generated sequence contains $T=120$ frames, corresponding to $4$ seconds of motion. 
We then calculated the mean and standard deviation of key performance metrics to assess their performance.

The results show that GCDance-Nash outperforms the baseline model EDGE by $30.27\%$ in FID\_hand and $55.58\%$ in FID\_body. 
Similarly, GCDance-Aligned improves over EDGE by $28.83\%$ in FID\_hand and $57.97\%$ in FID\_body. 
In terms of physical plausibility, GCDance-Nash and GCDance-Aligned achieve PFC scores of $0.13 \pm 0.001$ and $0.15 \pm 0.001$, which are close to the previous best FineNet, and both variants obtain the best PBC scores among all compared methods.
Regarding the BAS score, our models are slightly lower than that of LODGE by $0.0111$. Nevertheless, our model strikes a better balance between motion quality and diversity, leading to a more robust and generalizable performance.
Additionally, it is worth noting that DanceRevolution and MNET achieve significantly higher FID scores, which we attribute to discontinuities in their generated motions. 
Furthermore, DanceRevolution often produces repeated or frozen frames, resulting in low diversity scores. 
In contrast, MNET tends to generate overly jittery motions, leading to abnormally high diversity metrics that do not correspond to realistic movement quality. 
Bailando demonstrates state-of-the-art performance on the $24$-joint AIST++ dataset, as shown in Table~\ref{table_3}, but its performance degrades when evaluated on the higher-resolution $52$-joint FineDance dataset. 
This may be attributed to its design of the model, which directly predicts joint positions instead of rotations~\cite{siyao2022bailando,yang2023longdancediff}, potentially reducing accuracy when modeling more fine-grained skeletal structures.

\begin{table}[!t]  
\caption{A comparison on the AIST++ Dataset.%
\label{table_3}}
\centering
\resizebox{\columnwidth}{!}{
\begin{tabular}{cccccc}
\toprule
& \multicolumn{2}{c}{\textbf{Motion Quality}} & \multicolumn{2}{c}{\textbf{Motion Diversity}} & \multirow{2}{*}{\textbf{BAS}$\uparrow$} \\
\cmidrule(lr{0.35em}){2-3} \cmidrule(lr{0.35em}){4-5}
& FID\_k$\downarrow$ & FID\_m$\downarrow$ & Div\_k$\uparrow$ & Div\_m$\uparrow$ & \\
\midrule
FACT~\cite{li2021ai}            & 86.43  & 43.46  & 6.85  & 3.32  & 0.1607 \\
DanceNet~\cite{zhuang2020music2dance} & 69.18  & 25.49  & 2.86  & 2.85  & 0.1430 \\
Bailando~\cite{siyao2022bailando}     & 28.16  & \textbf{9.62}  & \textbf{7.83} & \underline{6.34} & 0.2332 \\
DiffDance~\cite{qi2023diffdance}      & \textbf{24.09} & 20.68  & 6.02  & 2.89  & \underline{0.2418} \\
EDGE~\cite{tseng2023edge}             & 42.16  & 22.12  & 3.96  & 4.61  & 0.2334 \\
LODGE~\cite{li2024lodge}              & 37.09  & 18.79  & \underline{5.58} & 4.85  & \textbf{0.2423} \\
\midrule
GCDance (Aligned)                     & 35.91  & 19.19  & 5.07  & 5.70  & 0.2321 \\
GCDance (Nash)                        & \underline{30.93} & \underline{18.25} & 5.22  & \textbf{6.71} & 0.2354 \\
\bottomrule
\end{tabular}
}
\vspace{-0.5cm}
\end{table}

Additionally, we trained our method on the publicly available AIST++ dataset, as shown in Table~\ref{table_3}. Following~\cite{li2021ai}, we utilized the FID\_m and Div\_m metrics, which evaluate the distributional spread of generated body part dances within the geometric feature space~\cite{muller2005efficient}. 
Since~\cite{muller2005efficient} does not provide geometric information for hand skeletons, it cannot be applied to the FineDance dataset. 
Due to the absence of genre information and hand motion data in the AIST++ dataset, our model does not achieve the best results. Nevertheless, GCDance shows improvements in multiple metrics compared to the baseline model, EDGE.

\subsection{Qualitative Results}
\begin{figure}[!t]
  \centering
  \includegraphics[width=.49\textwidth]{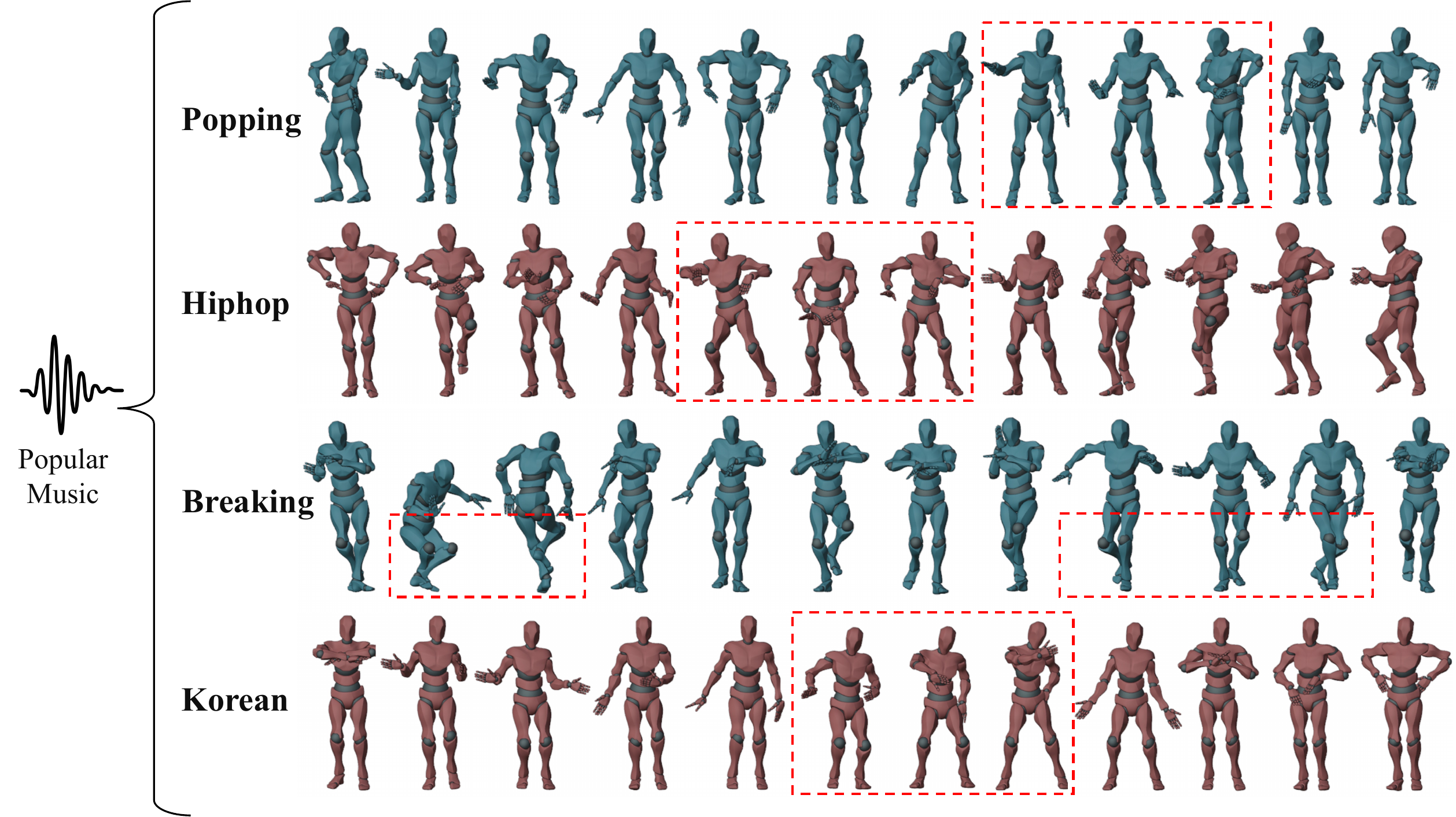}
  \caption{\textbf{Same music, different popular dance.} Boxed hand, leg, and full‑body poses highlight the salient stylistic features that distinguish each genre.}
  \vspace{-0.3cm}
  \label{fig_5}
\end{figure}

\begin{figure}[!t]
  \centering
  \includegraphics[width=.49\textwidth]{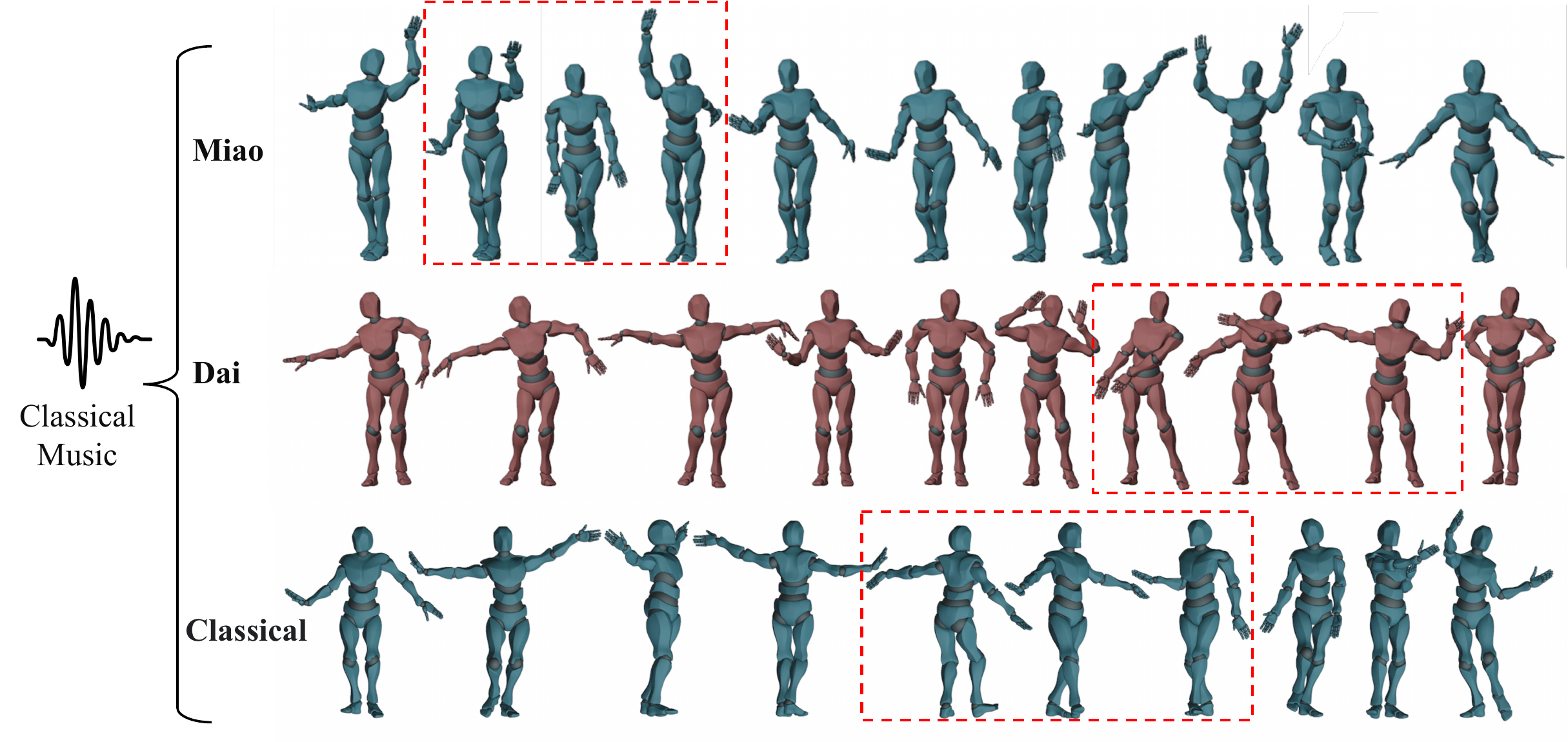}
  \caption{\textbf{Same music, different classical dance.} Boxed hand, leg, and full‑body poses highlight the salient stylistic features that distinguish each genre.}
  \vspace{-0.5cm}
  \label{fig_6}
\end{figure}

To verify the controllability of our model in generating dances of specific genres, we conducted experiments using the same pieces of music with different genre labels. 
We input a segment of popular modern music and provided four different genre labels: Popping, Hip-hop, Breaking, and Korean. 
Then we visualized the generated dance sequences as shown in Figure~\ref{fig_5}. 
Similarly, in Figure~\ref{fig_6}, we input the same piece of classical music but applied three different genre labels: Miao, Dai, and Classical dance. 
In the first set of results, the generated Popping sequence features sharp hits with smooth transitional waves, Hip-Hop features abundant arm movements complemented by small rhythmic hops. Breaking emphasizes dynamic footwork, and Korean dance reproduces iconic K-pop elements. 
In the second set, Miao folk dance displays interlacing arm swings, Dai folk dance exhibits fluid and seamless movements, and Classical dance highlights broad arm gestures with graceful turns. These results demonstrate the controllability of GCDance in producing diverse stylistic performances from the same musical input. Additional videos are available on our project page. Additional demonstration videos are available on our project page.

\begin{figure}[!t]
  \centering
  \includegraphics[width=.48\textwidth]{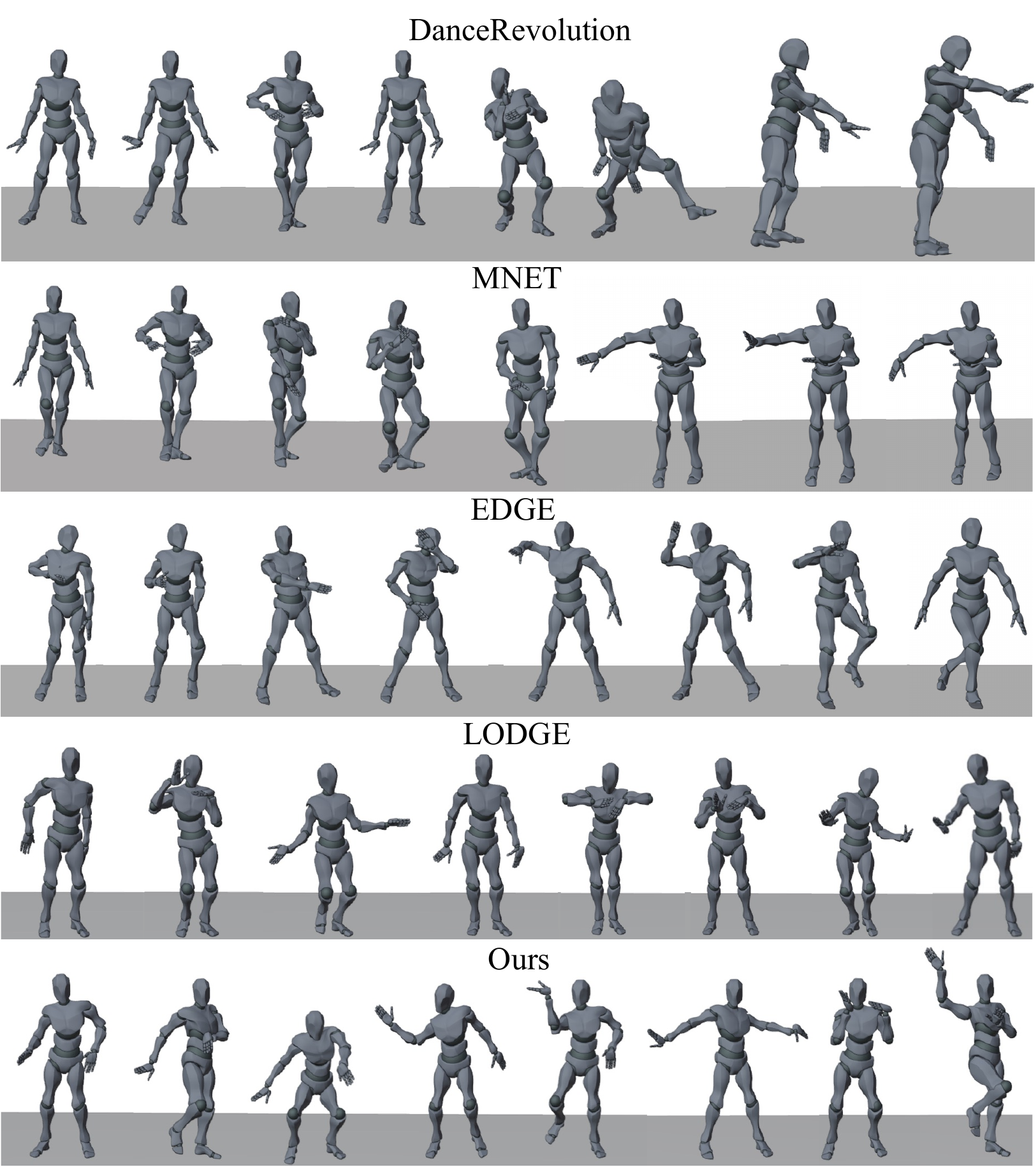}
  \caption{Visualization comparison of SOTAs methods.}
  \label{fig_4-5}
  \vspace{-0.5cm}
\end{figure}

Figure \ref{fig_4-5} presents a qualitative comparison between our method and four baselines.
DanceRevolution and MNET both suffer from motion stagnation after only a few seconds, reflecting poor temporal continuity and limited expressiveness. EDGE alleviates this freeze but introduces conspicuous artifacts, most notably unnatural hand trajectories and noticeable foot sliding. LODGE produces smoother kinematics yet  offers reduced stylistic diversity. By contrast, our approach delivers motions with higher perceptual fidelity, richer stylistic variation, and  coherence throughout the entire sequence.

\begin{table}[!t]
\caption{Ablation study. We evaluate the contribution of Foundation Model features (FM), the Genre Classification Module (GCM), and different Multi-Task Learning (MTL) strategies. Variants without MTL use fixed loss weights following~\cite{tseng2023edge}.%
\label{table_4}}
\centering
\footnotesize
\resizebox{\columnwidth}{!}{
\begin{tabular}{lcccccc}
\toprule
& \multicolumn{2}{c}{\textbf{Motion Quality}} & \multicolumn{2}{c}{\textbf{Motion Diversity}} & \multirow{2}{*}{\textbf{PFC}$\downarrow$} & \multirow{2}{*}{\textbf{BAS}$\uparrow$} \\  
\cmidrule(lr{0.35em}){2-3} \cmidrule(lr{0.35em}){4-5}
 & FID\_h$\downarrow$ & FID\_b$\downarrow$ & Div\_h$\uparrow$ & Div\_b$\uparrow$ & & \\
\midrule
w/ FM    &           23.70 &                      25.84&                      7.60&                      6.85&                  0.17&     0.2160\\
+ FM &    18.48            & 22.61            & 8.77          & 6.77               &  0.17       & 0.2188                \\
+ GCM&                  16.95&                 30.31&                   8.92&                  6.36&                      0.15&                      0.2170\\
\midrule
+ Aligned-MTL&    17.69 & 22.90 & 9.47 & 6.39 &  0.13 & 0.2238  \\
+ Nash-MTL&    18.06 & 21.67 & 9.01 & 6.84 &  0.13 & 0.2205   \\
\bottomrule
\end{tabular}
}
\vspace{-0.5cm}
\end{table}

\subsection{Ablation Study}
In Table~\ref{table_4}, we presented the ablation results by analyzing the effects of music feature composition, genre classification module, and multi-objective optimization strategy. 
It is shown that incorporating features from the music foundation model leads to consistent improvements across most metrics, highlighting the advantage of leveraging high-level semantic audio representations. Although the addition of the genre classification module improves controllability, it introduces an imbalance across different metrics by enhancing FID\_hand and PFC, while degrading FID\_body and Div\_body. With our multi-objective optimization strategy, the model achieves a more balanced performance and effectively improves the overall metric scores.

To better understand the impact of different music features on dance quality, we evaluated results generated using music features extracted by using various music foundation models, including CLAP~\cite{wu2023large}, Wav2Vec2.0~\cite{baevski2020wav2vec}, Jukebox~\cite{dhariwal2020jukebox}, and Wav2CLIP~\cite{wu2022wav2clip}, as well as hand-crafted features, including MFCC, STFT, and the 35-dimensional feature set provided by the FineDance dataset~\cite{li2023finedance}.
To fairly evaluate the impact of different music features, we used the backbone of our model without text classification and multi-task learning enhancements. 
This allowed us to isolate the effect of music feature design from other factors. 
Table~\ref{table_5} shows that our method, which incorporates music features extracted from a music foundation model and hand-crafted features, achieves the best overall performance. 
It demonstrates robust improvements in motion quality, diversity, and rhythm consistency over other music feature-based methods.
\vspace{-0.3cm}

\begin{table}[!]
\caption{Impact of Music Features on Generated Dance Quality%
\label{table_5}}
\centering
\footnotesize
\resizebox{0.5\textwidth}{!}{
\begin{tabular}{lccccccc}
\toprule
& \multicolumn{2}{c}{\textbf{Motion Quality}} & \multicolumn{2}{c}{\textbf{Motion Diversity}} &\multirow{2}{*}{\textbf{PFC}$\downarrow$}  &\multirow{2}{*}{\textbf{PBC}$\rightarrow$  }& \multirow{2}{*}{\textbf{BAS}$\uparrow$} \\  
\cmidrule(lr{0.35em}){2-3} \cmidrule(lr{0.35em}){4-5}
 & FID\_h$\downarrow$ & FID\_b$\downarrow$ & Div\_h$\uparrow$ & Div\_b$\uparrow$ & && \\
 \midrule
 GT & /& /& 11.8156& 10.1810& /&5.23&0.2318\\
\midrule
CLAP~\cite{wu2023large} & 29.64 & 27.52 & 8.11 & 6.10 &0.23&3.43& 0.2076 \\
Wav2Vec2.0~\cite{baevski2020wav2vec} & 21.78 & 34.65 & 8.61 & 6.32 &0.20&3.93& 0.2026 \\
Jukebox~\cite{dhariwal2020jukebox} & 23.02 & 32.26 & 7.41 & 6.38 &0.24&3.35& 0.2238 \\
Wav2CLIP~\cite{wu2022wav2clip} & 22.19 & 33.65 & 8.51 & \textbf{8.85} &0.17&4.40& \textbf{0.2276} \\
STFT & 23.70 & 25.84 & 7.60 & 6.85 &0.16&3.86& 0.2160 \\
MFCC & 27.63 & 33.74 & 8.42 & 8.50 &0.19&3.81& 0.2123 \\
35-Feature Group*~\cite{li2023finedance} & 20.61 & 25.41 & 8.22 & 5.84 &\textbf{0.15}&3.32& 0.2028 \\
\midrule
Wav2CLIP+STFT (Ours) & \textbf{18.48} & \textbf{22.61} & \textbf{8.77} & 6.77 &0.17& \textbf{4.43} & 0.2188 \\
\bottomrule
\end{tabular}
}
\vspace{-0.3cm}
\end{table}

\subsection{User study}
We conducted a user study involving $20$ participants at ANONYMIZED to evaluate the quality and controllability of the dance motions generated.
For each method, $270$ music–dance pairs were generated on the FineDance test set. From these, we randomly selected the same $8$ pairs across all methods to ensure a fair comparison.

For motion quality evaluation, participants rated each video on overall quality, smoothness, and synchronization with the music rhythm. As shown in Table~\ref{tab-control}, our method consistently outperforms all baselines, achieving at least a $63.26\%$ higher preference rate. For controllability evaluation, participants were presented with $8$ pairs of dance videos sharing the same genre label, one generated with genre control and the other using the ground truth label. They were asked to choose the video that better matched the given genre description. The results show that our generated dances were selected almost as frequently as the ground truth, confirming the strong genre controllability and consistency of GCDance.
\vspace{-0.3cm}

\begin{table}[!t]
\caption{User Study on Generated Dance Samples.%
\label{tab-control}}
\centering
\footnotesize
\setlength{\tabcolsep}{0.77pt}
\resizebox{0.48\textwidth}{!}{
\begin{tabular}{lccccc}
\toprule
\textbf{Model}           & GCDance-Aligned & GCDance-Nash & FineDance & EDGE & Bailando \\ 
\midrule
\textbf{Wins}           & /       & 53.57\%       & 63.26\%         & 78.51\%    & 89.28\%        \\
\textbf{Control Score} & 46.87\%    &  45.83\%
& /         & /    & /        \\
\bottomrule
\end{tabular}
}
\vspace{-0.5cm}
\end{table}

\subsection{Model Efficiency}
We also compared the efficiency of different models in Table~\ref{table_6}. 
During the inference phase, we evaluated the model parameters and inference times for generating $4$-second dance sequences. 
The table shows that our model achieves the inference time of $0.22$ seconds with $87M$ parameters, achieving a good balance between model size and speed. 
It matches FineNet in speed with fewer parameters and substantially outperforms Bailando, FACT, and LODGE in both efficiency and computational cost. 
Although slightly slower than EDGE, which attains the fastest inference time with fewer parameters, GCDance offers a better overall balance between efficiency and motion generation quality.

\begin{table}[!t]
\caption{Model parameters and \textit{per-instance} inference time.%
\label{table_6}}
\centering
\renewcommand{\arraystretch}{0.85} 
\footnotesize
\begin{tabular}{lcc}
\toprule
\textbf{Model} & \textbf{Parameters} & \textbf{Inference Time} \\
\midrule
FACT~\cite{li2021ai}            & 120M & 33.20s \\
Bailando~\cite{siyao2022bailando} & 152M & 0.94s \\
EDGE~\cite{tseng2023edge}       & 50M  & 0.13s \\
FineNet~\cite{li2023finedance}  & 94M  & 0.23s \\
LODGE~\cite{li2024lodge}        & 236.8M  & 1.89s\\
\midrule
GCDance (Ours)                  & 88M  & 0.22s \\
\bottomrule
\end{tabular}
\vspace{-0.5cm}
\end{table}

\vspace{-0.2cm}
\section{Conclusion}
In this paper, we presented GCDance, a diffusion-based 3D dance generation framework conditioned on both music and text prompts. 
By incorporating a genre classification module and leveraging features from a pretrained music foundation model, our method enabled precise and controllable synthesis of genre-consistent dance motions while preserving high motion quality and diversity. 
Furthermore, we used a multi-objective optimization strategy to balance the competing objectives, such as spatial accuracy, physical plausibility, and genre alignment, used for network training. 
Extensive experimental results obtained on the FineDance and AIST++ datasets demonstrated the superiority of our method over the existing approaches both qualitatively and quantitatively.

However, the proposed GCDance method focuses on genre-level control of the generated dance sequences.
It lacks the capability for fine-grained manipulation of specific motion attributes. 
In the future, we will further improve the model to enable fine-grained local editing for dance motion generation and varying the input text prompts at each decoding time step. 

\begingroup
 
\bibliographystyle{IEEEtran}
\bibliography{Reference}
\endgroup


\newpage

\end{document}